\documentclass{aa}  

\usepackage{graphicx}
\usepackage[colorlinks=true,citecolor=blue]{hyperref}
\usepackage{txfonts}
\begin{document} 

   \title{Variation on a Zernike wavefront sensor theme: optimal use of photons}

   \author{V. Chambouleyron
          \inst{1}\fnmsep\inst{2}
          \and
          O. Fauvarque\inst{3}
          \and
          J-F. Sauvage\inst{2}\fnmsep\inst{1}
          \and
          K. Dohlen\inst{1}
          \and
          N. Levraud
          \inst{2}\fnmsep\inst{1}
          \and
          A. Vigan\inst{1}
          \and
          M. N'Diaye\inst{4}
          \and
          B. Neichel\inst{1}
          \and
          T. Fusco\inst{2}\fnmsep\inst{1}
          }

   \institute{Aix Marseille Univ, CNRS, CNES, LAM, Marseille, France\\
              \email{vincent.chambouleyron@lam.fr}
         \and
             DOTA, ONERA, Université Paris Saclay, F-91123 Palaiseau, France
          \and
          IFREMER, Laboratoire Detection, Capteurs et Mesures (LDCM), Centre Bretagne, ZI de la Pointe du Diable, CS 10070, 29280, Plouzane, France
          \and
          Université Côte d’Azur, Observatoire de la Côte d’Azur, CNRS, Laboratoire Lagrange, France
             }

 
  \abstract
  {} 
  {The Zernike wavefront sensor (ZWFS) is a  concept belonging to the wide class Fourier-filtering wavefront sensor (FFWFS). The ZWFS is known for its extremely high sensitivity while having a low dynamic range, which makes it a unique sensor for second stage adaptive optics (AO) systems or quasi-static aberrations calibration sensor. This sensor is composed of a focal plane mask made of a phase shifting dot fully described by two parameters: its diameter and depth. In this letter, we aim to improve the performance of this sensor by changing the diameter of its phase shifting dot.} 
  {We begin with a general theoretical framework providing an analytical description of the FFWFS properties, then we predict the expected ZWFS sensitivity for different configurations of dot diameters and depths. The analytical predictions are then validated with end-to-end simulations. From this, we propose a variation of the classical ZWFS shape which exhibits extremely appealing properties.}
  {We show that the ZWFS sensitivity can be optimized by modifying the dot diameter and even reach the optimal theoretical limit, with a trade-off for low spatial frequencies sensitivity. As an example, we show that a ZWFS with a $2\ \lambda/D$ dot diameter (where $\lambda$ is the sensing wavelength and $D$ the telescope diameter), hereafter called  Z2WFS, exhibits a sensitivity twice higher than the classical $1.06\ \lambda/D$ ZWFS for all the phase spatial components except for tip-tilt modes. Furthermore, this gain in sensitivity does not impact the dynamic range of the sensor, and the Z2WFS exhibits a similar dynamical range as the classical $1.06\ \lambda/D$ ZWFS. This study opens the path to the conception of diameter-optimized ZWFS.}
   {}


   \maketitle
%

\section{Introduction}

The role of a wavefront sensor (WFS) is to encode the phase information at the entrance of an optical system into intensities on a detector. For ground-based astronomy, WFS are mostly used for Active or Adaptive Optics (AO), in conjunction with a wave-front control strategy in order to compensate for optical aberrations induced by the atmosphere or the telescope itself. In the context of astronomy, one of the main driver for a WFS design is its sensitivity, or in other words, its ability to provide an accurate measurement in the presence of noise. Sensitivity is therefore a useful metric to assess WFS performance in terms of photon noise, which is related to key quantities in AO field: loop speed and sky-coverage. Existing WFS can be separated into two main categories, usually defined as focal plane WFS, and for which the measurements are done in a focal plane (like the Shack-Hartmann wavefront sensor) and pupil plane WFS, for which the measurements are done in a pupil plane. Among this later category, the Fourier Filtering WFS (FFWFS) represents a wide class of sensors of particular interest thanks to their superior sensitivity. From a general point of view, a FFWFS consists of a phase mask located in an intermediate focal plane which performs an optical Fourier filtering. As such, the Zernike phase mask \citep{zernike1934,bloemhof,Kjetil2006,wallace2011} forms a FFWFS, hereafter called Zernike WFS (ZWFS). In this case, the filtering element is a phase shifting dot which is, for a given substrate, fully described by two parameters: its diameter and its depth (or phase-shift). In a classical implementation, the ZWFS phase dot has a diameter of $1.06\ \lambda/D$ (where $\lambda$ is the sensing wavelength and $D$ the telescope diameter) and a phase shift of $\pi/2$. This ZWFS is known to be one of the most sensitive WFS \citep{Guyon_2005}. Its drawback being its limited dynamic range, it has therefore been mostly implemented as a second-stage WFS, or as a quasi-static aberrations calibration sensor like on VLT/SPHERE \citep{ZeldaMamadou, viganSphere}. In this paper, we show that the classical implementation of the ZWFS with a phase dot diameter of $1.06\ \lambda/D$ is actually not optimal: by using a larger dot diameter, the sensitivity of the ZWFS can be significantly improved at the expense of lower spatial frequencies sensitivity, and even reach a performance close to the theoretical limit. For this reason, Section 2 starts from a theoretical study of the ZWFS based on a general convolutional formalism for FFWFS \citep{fauvOptica}. This analytical work shows that the sensitivity of the ZWFS can be improved by increasing its dot diameter. We then confirm the theoretical results with end-to-end simulations in Section 3 and we show that a gain in sensitivity by a factor two can be reached without impacting the dynamic range of the sensor. Conclusions are given in Section 4.

\section{Theoretical analysis of the ZWFS sensitivity with a convolutional approach}

\subsection{Definition of a FFWFS sensitivity}

Following the formalism introduced by \cite{fauvOptica}, the raw intensities recorded by a FFWFS are processed with a return-to-reference operation. It simply consists in removing from the FFWFS recorded intensities map $I(\phi)$ the one corresponding to the reference phase $I_{0}$ (usually a flat wavefront). We also posit that all the intensities are normalized by the number of photons. The resulting quantity is called tared intensities:

\begin{equation}
\Delta I(\phi) = I(\phi)-I_{0}
\end{equation}

One of the most important performance criteria for a sensor is its behaviour in terms of noise propagation. This criteria is encoded in a quantity called sensitivity which depends on the energy in the columns of the Interaction Matrix (IM - \cite{1992Rigaut}). For a given WFS, each column of the IM is built as the linear response of the sensor to a given mode $\phi_{i}$ which is usually obtained experimentally through a "push-pull" method:

\begin{equation}
\delta I(\phi_{i}) = \frac{I(\epsilon\phi_{i})-I(-\epsilon\phi_{i})}{2\epsilon}
\end{equation}

where $\epsilon$ is the amplitude of the mode. The sensitivity $s$ for a given mode $\phi_{i}$ is then defined through the Euclidean norm:

\begin{equation}
s(\phi_{i}) = \frac{||\delta I(\phi_{i})||_{2}}{||\phi_{i}||_{2}}\,.
\label{eq:sensitivity}
\end{equation}

For a given uniform noise distribution $\sigma_{n}$, the noise propagation coefficient $\sigma_{WFS}$ for a mode $\phi_{i}$ is then related to the sensitivity by the following relationship:

\begin{equation}
\sigma^{2}_{WFS} = \sigma^{2}_{n}s(\phi_{i})^{-2}\,.
\end{equation}

For ground-based astronomy, where WFS are usually implemented within an AO loop, the sensitivity is a critical metric as it describes how the system performs in the presence of noise. Optimising the WFS sensitivity has always been one of the main motivation in the conception of new WFS.

Finally, it should be noticed that we choose to visualize the FFWFS sensitivity as a 2 dimensional map along the spatial frequencies of the wavefront. It consists in calculating sensitivity with respects to the Fourier modes $\phi_{i}$ (close from what was done in \cite{jensen-clem}) which are simply defined by the sum of a cosine and a sine carrying a given spatial frequency $\textbf{f}$. The following quantity then encodes the sensitivity:
\begin{equation}
s_{\textbf{f}} = \sqrt{s\Big(\cos_{\textbf{f}}\Big)^{2}+s\Big(\sin_{\textbf{f}}\Big)^{2}}\,.
\label{eq:e2e_fourier}
\end{equation}

\subsection{A convolutional approach to compute FFWFS sensitivity}

The FFWFS sensitivity can be computed based on a convolutional model, as described in \cite{fauvConv}. This model assumes that the sensor can be fully characterized by an impulse response $\textbf{IR}$ that links the entrance phase to the measured tared intensities: 

\begin{equation}
\Delta I(\phi) \approx \textbf{IR}\star\phi\,,
\end{equation}


where $\star$ stands for the classical convolutional product. A convenient aspect of the convolutional approach is the fact that one can compute the transfer function $\textbf{TF}$ of a FFWFS. FFWFS can be described by two parameters: their phase masks $m$, and their weighting functions $\omega$ which describe the energy distribution in the focal plane during one acquisition time of the sensor. We precise that this function is normalized to 1 in order to ensure energy conservation. Assuming that $\omega$ is a real function and that $\omega$ and $m$ are both centro-symmetric, which is generally the case for most of the known FFWFS, $\textbf{TF}$ is expressed through the simple following formula:

\begin{equation}
\textbf{TF}=2\text{Im}\big[m\star\overline{m\omega}\big]\,,
\label{eq:TF}
\end{equation}
where $\text{Im}$ is the imaginary part and the bar is the complex conjugate operator. From the knowledge of a FFWFS transfer function, it is then possible to compute the sensitivity with respect to spatial frequencies thanks to the formula exposed in \cite{fauvConv}:

\begin{equation}
s_{\textbf{f}} \approx \sqrt{|\textbf{TF}|^2\star \textbf{PSF}}\Big|_{\textbf{f}}\,,
\label{eq:sens}
\end{equation}
where the quantity \textbf{PSF} is the Point Spread Function of the system. Its energy correspond to the incoming flux and is normalized to 1.  


At this point, it is important to note that the sensitivity is bounded. Since the mask transmission,  $|m|$, cannot be greater than one ($|m| \leq 1$), Equation \ref{eq:TF} implies that $\textbf{TF}|_{\textbf{f}}\leq 2$. Hence, given Equation \ref{eq:sens}, we conclude that, in the frame of our normalizations, the sensitivity cannot be greater than 2:

\begin{equation}
\forall \textbf{f},\ s_{\textbf{f}}  \leq 2\,.
\label{eq:TFmax}
\end{equation}

This is an important result as it defines the theoretical limit for a FFWFS sensitivity.

\subsection{Application to the ZWFS}

The convolutional formalism introduced in the previous section is now applied to the ZWFS in order to find a simple formula of its sensitivity according to the mask parameters. As it was previously mentioned, the ZWFS mask is defined by two free parameters:
\begin{itemize}
    \item[$\bullet$] The depth (phase shift) of the dot $\delta$. For the classical ZWFS: $\delta = \pi/2$,
    \item[$\bullet$] Its diameter $p$. For the classical ZWFS: $p = 1.06\ \lambda/D$.
\end{itemize}

\noindent As described in \citet{ZWFSmamadou}, the dot diameter value was chosen in order to get an equivalent flux inside and outside the focal plane dot. This configuration with $p = 1.06\ \lambda/D$ also allows to get an uniform reference intensity distribution. The purpose of this section is to demonstrate that this choice of $p = 1.06\ \lambda/D$ is actually not optimal, and that the sensitivity can be improved with a larger dot. 

\begin{figure}[ht]
\centering
\includegraphics[scale=0.45]{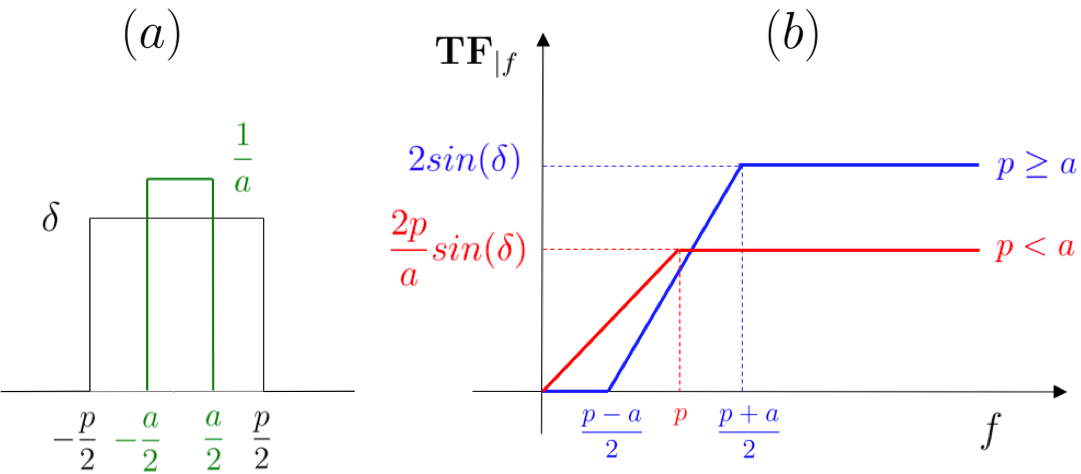}
\caption{\textbf{(a)} : Simplified 1D framework for convolutional derivations: in black, the dot diameter equals to $p$ and the phase shift equals to $\delta$. In green, the PSF is approximated by a normalized top-hat function with a diameter $a$. \textbf{(b)} : Transfer function of the ZWFS for two cases in red, the dot size is smaller than the PSF. In blue, the dot is larger than the PSF. The optimal case appears for $p=a$ and not $p = a/2$ as it is done for the classical ZWFS.}
\label{fig:fig1}
\end{figure}

For the sake of clarity, we carry this study in one dimension. As such, the spatial frequency vector $\textbf{f}$ becomes the scalar frequency $f$. We further assume that the weighting function $\omega$, which corresponds to the PSF, can be described as a top-hat function of diameter radius $a$. This simplified geometry is summarized by Figure \ref{fig:fig1}-(a). From this simplified geometry, we then compute $\textbf{TF}|_{f}$ and plot this quantity in Figure \ref{fig:fig1}-(b). Note that the full derivation can be found in appendix \ref{appendixA}.

From Figure \ref{fig:fig1}-(b), one can distinguish two cases: 
\begin{itemize}
    \item[$\bullet$] \textbf{$p \geq a$}, i.e. the dot diameter is larger than the PSF characteristic size. Frequencies above $(p+a)/2$ reach the sensitivity $2\sin(\delta)$. Frequencies below $(p-a)/2$ are at 0.
    \item[$\bullet$] \textbf{$p < a$}: the dot diameter is smaller than the PSF size. Frequencies over $p$ have a value of $2p/a\times sin(\delta)$, which is smaller than the theoretical limit of $2$.
\end{itemize}

From this simplified model one can first conclude that a phase shift of $\delta = \pi/2$ will maximize the sensitivity as expected. But surprisingly, it shows that for this phase shift, a dot radius of $p=a$ offers a sensitivity of the optimal value for almost all the modes. At this point it is important to remember that the value of $\textbf{TF}|_{f}$ directly sets the sensor sensitivity through Equation \ref{eq:sens}. It is therefore possible to design a ZWFS that reaches the theoretical sensitivity value. As a comparison, the classical ZWFS configuration \citep{ZWFSmamadou} uses $p = a/2$, which leads to a sub-optimal \textbf{TF} value of $1$ for frequencies above $p$.

Although this simplified study uses strong assumptions, it shows that the ZWFS can be further optimized compared to its classical form. In the next section, we demonstrate that these simplified results are actually accurate and enable us to build the most sensitive sensor ever proposed.

\section{Towards an optimal ZWFS}

Following the results from the convolutional approach, the goal of this section is to make use of numerical simulations to confirm the sensitivity of the ZWFS with respect to its dot diameter, and eventually to propose an optimal configuration. For that, we consider different configurations for a dot diameter ranging from 0 to 5 $\lambda/D$. The phase shift is set to $\delta = \pi/2$ for the rest of this letter. 

\subsection{Impact of the dot diameter on the ZWFS Sensitivity}

\begin{figure}[htbp]
\centering
\includegraphics[scale=0.4]{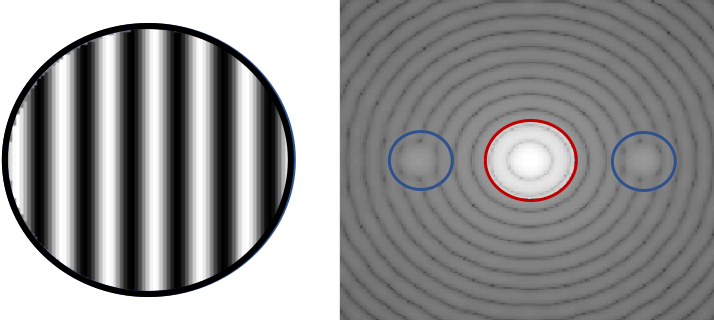}
\caption{Cosine phase (left insert, linear scale) corresponding to a spatial frequency of 6 cycles in D and its corresponding PSF speckles (right insert, logarithmic scale, in blue). This frequency lies outside of the dot footprint, here $5\ \lambda/D$ (in red).}
\label{fig:sinus_wave}
\end{figure}

\begin{figure}[htbp]
\centering
\includegraphics[scale=0.5]{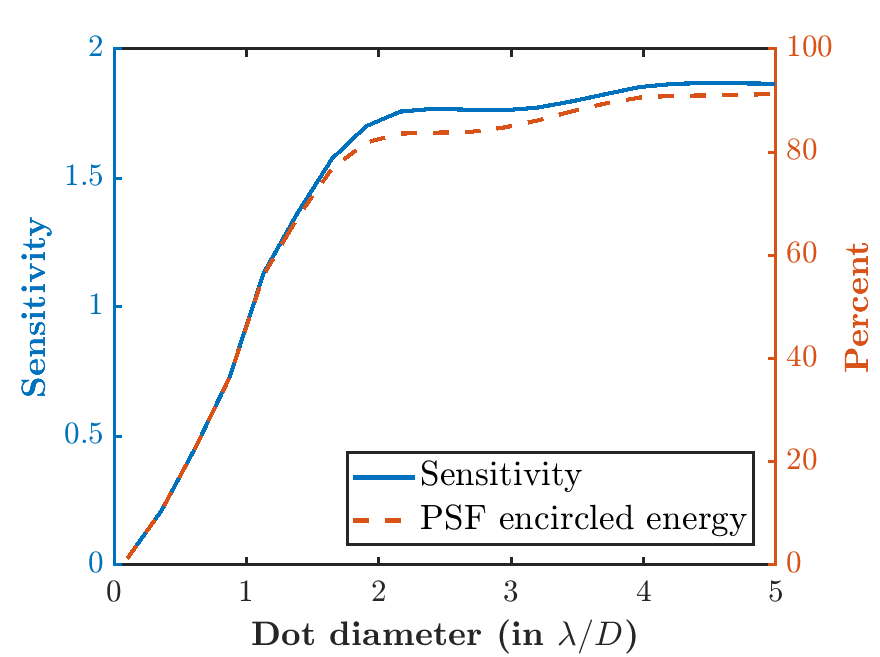}
\caption{Sensitivity evolution for a frequency outside of the dot, while increasing dot diameter. The sensitivity is in strong accordance with the proportion of the PSF energy located inside the dot, as predicted by the convolutional approach.}
\label{fig:diameter}
\end{figure}

As a first step, we want to illustrate the impact of the dot diameter on sensitivity for a spatial frequency located outside of the dot (horizontal part of the curves Figure \ref{fig:fig1}-(b)). For that purpose, we arbitrary choose a spatial frequency with 6 cycles over the pupil (left insert of Figure \ref{fig:sinus_wave}), which is far enough from the maximum diameter dot value ($5\ \lambda/D$ \textit{i.e} a radius of $2.5\ \lambda/D$). This configuration is illustrated in right insert of Figure \ref{fig:sinus_wave}. The sensitivity results for this spatial frequencies are shown in Figure \ref{fig:diameter}.

As predicted by the convolutional approach, Figure \ref{fig:diameter} shows that the sensitivity for a high spatial frequency increases with the dot diameter. This behaviour has been discussed briefly in previous literature \citep{Ruane_2020} without further analysis. It is here explained thanks to the convolutional model. It is also interesting to notice that the sensitivity growth is closely following the PSF encircled energy in the dot diameter, confirming the analytical results presented in Figure \ref{fig:fig1}. For a classical ZWFS, with a dot diameter of 1.06 $\lambda/D$, the sensitivity is actually far from being optimal.

\begin{figure}[ht]
\centering
\includegraphics[scale=0.55]{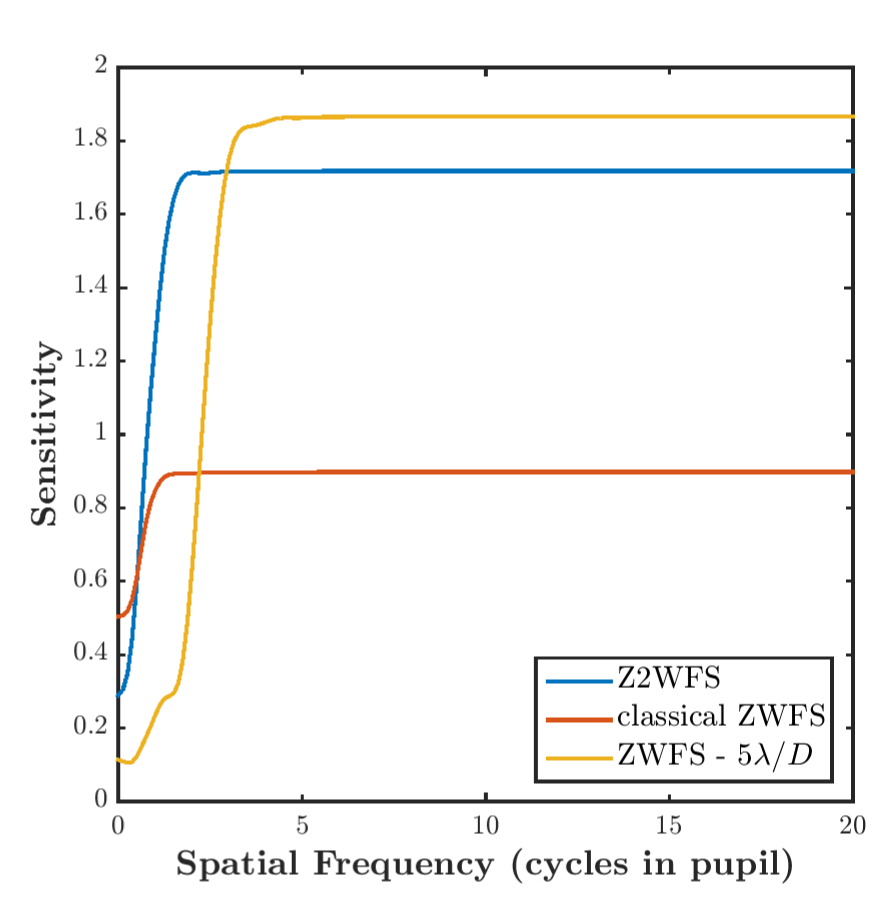}
\caption{Sensitivity curves for different dot diameters. This figure has to be compared with the convolutional approach Figure \ref{fig:fig1}.}
\label{fig:diamFreq}
\end{figure}

However one cannot just increase the dot diameter inconsiderately because the sensitivity to frequencies lying inside the dot would drop to zero (Figure \ref{fig:fig1}). We illustrate this effect in Figure \ref{fig:diamFreq}, where we plot the sensitivity curves with respect to a wide range of spatial frequencies for the three dot diameters configurations $p=1.06,~2$ and $5\lambda/D$. For $p = 5\lambda/D$, the sensitivity to high-spatial frequencies (those larger than 5 cycles per pupil) almost reaches the theoretical limit of 2, however the sensitivity becomes close to 0 for low-spatial frequencies (those smaller than 3 cycles per pupil). There is therefore a trade-off between enhanced sensitivity and unseen modes. In the following, we propose to choose the configuration with $p = 2\lambda/D$ and we call this specific configuration the Z2WFS. 

\begin{figure}[htbp]
\centering
\includegraphics[scale=0.4]{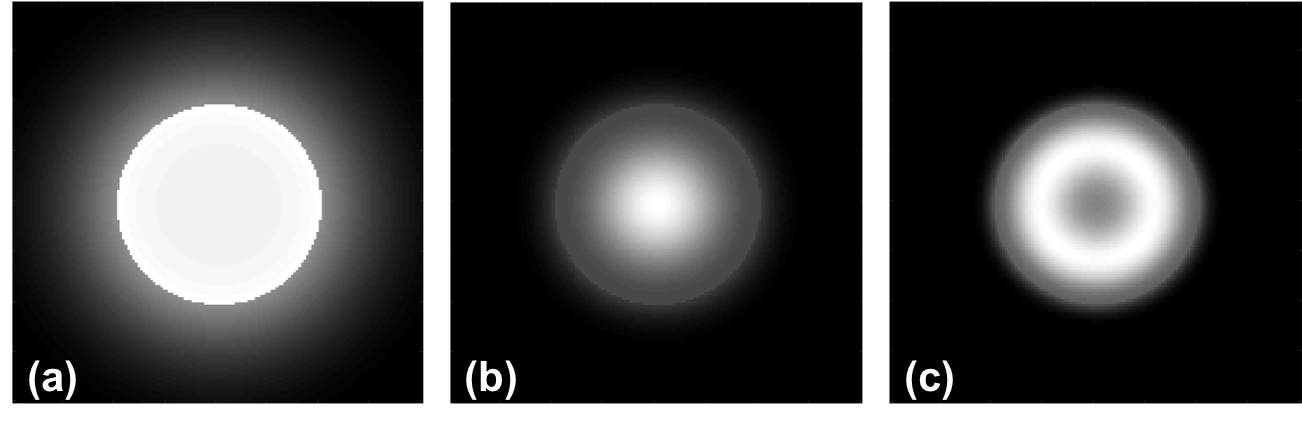}
\caption{Reference intensities of the ZWFS for different dot diameter. \textbf{(a)}: $p = 1.06\ \lambda/D$. \textbf{(b)}: $p = 2\ \lambda/D$. \textbf{(c)}: $p = 5\ \lambda/D$.}
\label{fig:Iref}
\end{figure}

As a remark, we also emphasize that the reference intensities, i.e. the intensity distribution for a flat wavefront, change with the value of $p$. This is illustrated in Figure \ref{fig:Iref} for the previous three different values of $p$. The classical ZWFS shows a flat reference illumination, while the Z2WFS appears to be less uniform. This spatial distribution could involve practical issues in terms of detector dynamics or for complex pupil shapes, as for instance central obscuration or spiders. These potential practical implementation issues are beyond the scope of this paper. In this letter, we will only assume a full aperture pupil with monochromatic light for the sake of clarity. It is to be noted that there is no sticking point here, the formalism and results developed here are maintained with a central obscuration in the pupil.  

\subsection{Comparison with other FFWFS}

\begin{figure}[htbp]
\centering
\includegraphics[scale=0.5]{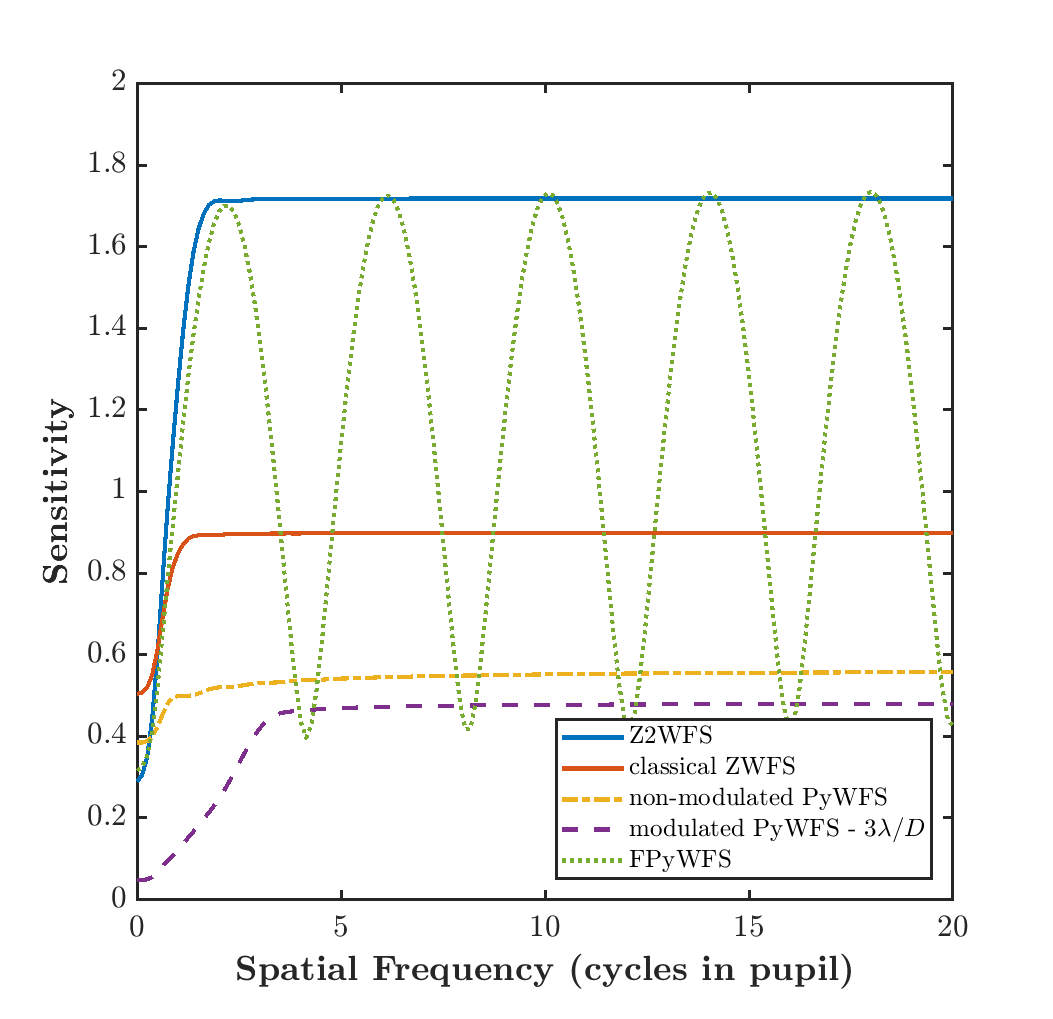}
\caption{Sensitivity curves for different FFWFS. We can distinguish the ZWFS class and the PyWFS class. The Z2WFS overtakes all the other sensors.}
\label{fig:comparisonFourier}
\end{figure}

In this section, we compare the Z2WFS with other well-know FFWFS: the classical ZWFS with $p = 1.06\ \lambda/D$, the non-modulated PyWFS (\cite{raga}), the modulated PyWFS (with here a modulation radius of $3\ \lambda/D$) and a flattened pyramid (FPyWFS) proposed by \cite{fauvFP} with a pupils overlapping rate of $75$\%. Spatial frequencies basis, i.e. Fourier basis, is chosen for this comparison. Results are given Figure \ref{fig:comparisonFourier}. First, we retrieve well-know results as for instance the gain around a factor $2$ in sensitivity between the classical ZWFS and the PyWFS.
We can also highlight the behaviour of the FPyWFS showing oscillating sensitivity and peaks for some specific frequencies, as described in \cite{fauvFP}. (The explanation of the PyWFS class behaviour through to the convolutional approach is also given in appendix \ref{appendixB}.) The Z2WFS is clearly the most sensitive sensor, except for extremely low frequencies. As expected from Figure \ref{fig:diameter}, it has a sensitivity twice better than the classical ZWFS for almost all modes and is four times more sensitivity than the non-modulated PyWFS.\\

\begin{figure}[htbp]
\centering
\includegraphics[scale=0.4]{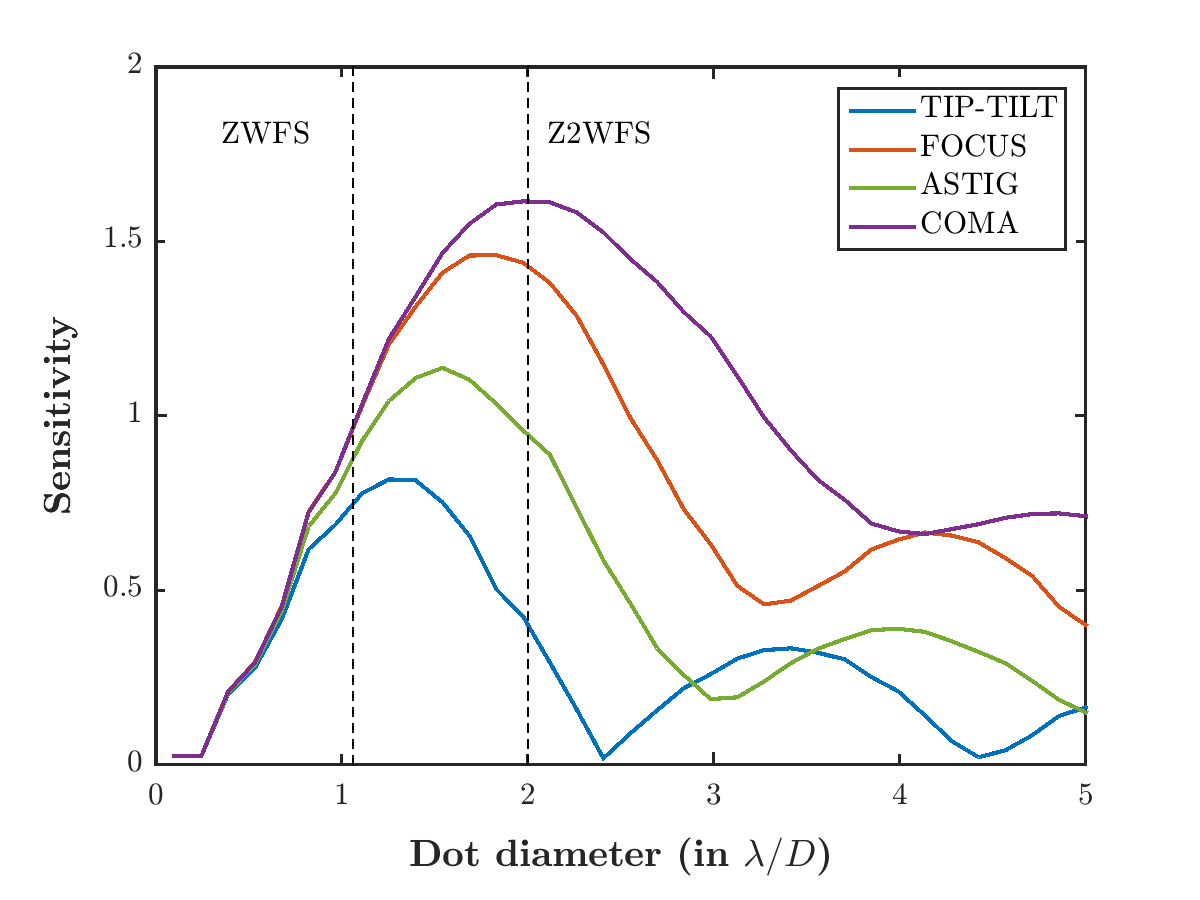}
\caption{Evolution of the low order Zernike modes sensitivities with respect to the dot diameter. We can see that the Z2WFS senstivity is lower for the tip-tilt modes, but higher for all the other ones. The classical ZWFS is not even optimized for the tip-tilt modes.}
\label{fig:comparisonZernike}
\end{figure}

The behavior at low spatial frequencies deserves some further analysis: we  plot  Figure \ref{fig:comparisonZernike} the sensitivity with respects to the tip-tilt and focus modes (which are the lowest frequency Zernike modes) for the ZWFS class with a dot diameter ranging from 0 to 5 $\lambda/D$. For the tip-tilt modes, the Z2WFS is twice less sensitive than the classical ZWFS, but the Z2WFS provides better results for the focus. Even if the Z2WFS is less sensitive for tip-tilt than the classical ZWFS, it is important to note that it remains as sensitive as the non-modulated PyWFS which is around 0.4. As a remark, it is interesting to see that the sensitivity curve for the tip-tilt is following the PSF shape: for the edge of the dot lying on a dark area of the PSF, the sensitivity drops to 0. 
By taking a Z1.5WFS ($p=1.5\ \lambda/D$), one could have a better sensitivity for all the frequencies compared to the classical ZWFS, but a lowest gain overall compared to Z2WFS.

To conclude this section, we demonstrated that a Z2WFS significantly improves the sensitivity, and approaches the ideal FFWFS behavior. When compared to the classical ZWFS, the gain in sensitivity for all modes, except the Tip-Tilt, is a factor 2. In the next section we investigate if this gain in sensitivity costs in dynamic range.

\subsection{Dynamic Range}

To complete our study, we now compare the dynamic range of the Z2WFS with the classical ZWFS. A drastic loss of dynamic range while increasing the dot diameter could indeed prevent from a practical utilisation of the Z2WFS. To calculate this quantity with respects to a given mode $\phi_{i}$, we evaluate its capture range $C_{\phi_i}$. To calculate it, we look at the lowest amplitude value (in absolute value) such that :
\begin{equation}
\left.\frac{\text{d}\Delta I(a\phi_{i})}{\text{d}a}\right|_{a_{0}} = 0\,.
\label{eq:capture}
\end{equation}
We then define the capture range as $C_{\phi_i} = 2a_{0}$, where the factor 2 allows one to take into account negative and positive amplitudes in the capture range calculation. 

The capture range can be larger than the pure linearity regime. However, we decided to use this definition for two reasons. First because it defines the amplitude below which we are ensured that a closed loop system will eventually converge. Indeed, even if the measurement is not linear anymore, there is still a one-to-one correspondence with the input signal. Secondly, because, the ZWFS measurements are often processed through non-linear reconstructors (\cite{ZWFSmamadou}, \cite{Steeves:20}) that can be perfectly applied to Z2WFS or other variations of the ZWFS.

\begin{figure}[ht]
\centering
\includegraphics[scale=0.5]{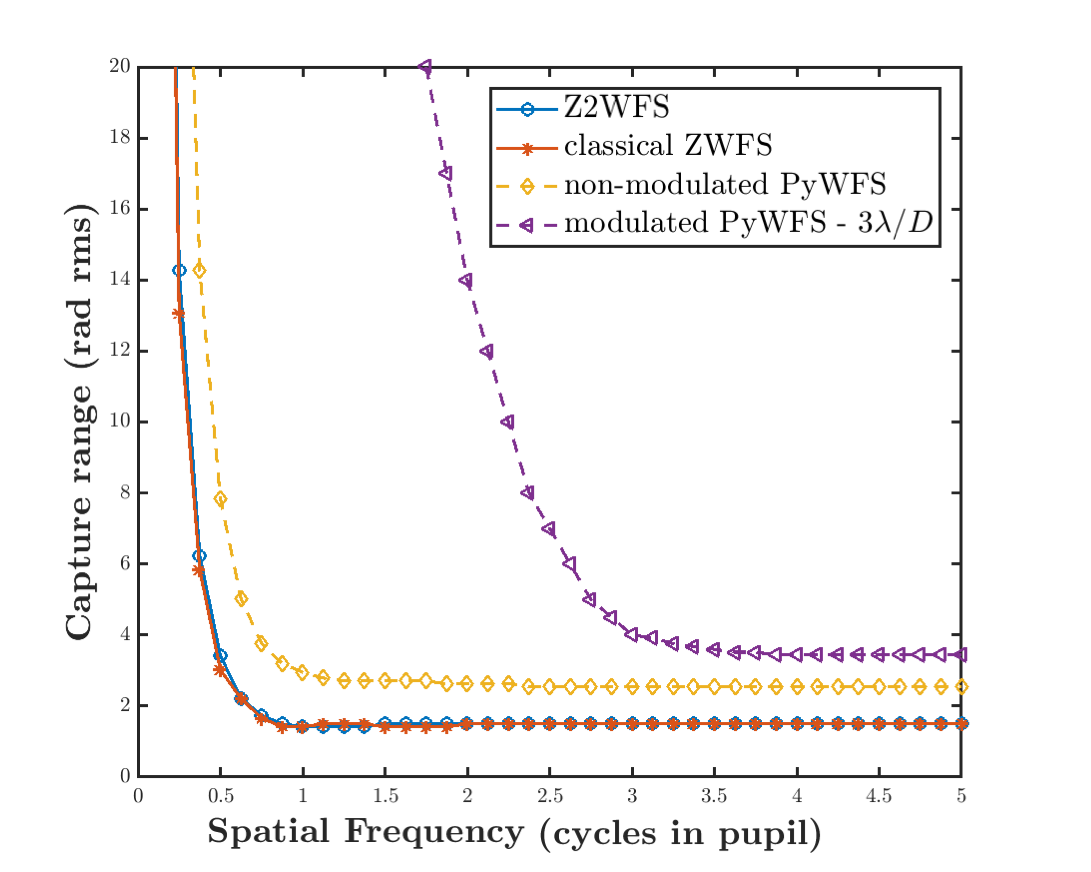}
\caption{Capture range of different FFWFS. The Z2WFS has the same capture range as the classical ZWFS.}
\label{fig:capture_range}
\end{figure}

Capture range values for cosine phase modes at frequencies ranging from $0$ to $5$ cycles in diameter are given Figure \ref{fig:capture_range} for Zernike and Pyramid wave-front sensors. The PyWFS class have a better capture range over all spatial frequencies, matching the fact that sensitivity and dynamic range are competing properties. This graph also confirms the great benefit in terms of dynamic range provided by the modulation of the PWFS. More importantly, we see that the Z2WFS exhibits the same capture range as the classical ZWFS for high frequencies and is even higher for the lowest frequencies where the Z2WFS sensitivity goes below the classical ZWFS one. The Z2WFS is therefore more sensitive than the classical ZWFS while exhibiting the same capture range.

\section{Conclusion}

In this paper, we provided a physical description of the sensitivity behaviour for the ZWFS class, and in particular we studied the sensitivity evolution for different dot diameters. We showed that it was possible to significantly improve the current sensitivity of the ZWFS at the expense of the lower spatial frequencies, simply by increasing the dot diameter. The resulting sensitivity can even almost reach the fundamental limit of FFWFS. We further studied the specific case of a dot diameter of $2\ \lambda/D$ - called Z2WFS, which exhibits a averaged gain of sensitivity by a factor two (with a loss of sensitivity compared to the classical ZWFS only for the tip-tilt modes), without loss of the dynamic range with respect to the classical ZWFS. This new sensor then becomes the most sensitive WFS available for ground-based astronomy. It still exhibits the low dynamic range of the ZWFS, but as for the PyWFS, modulation schemes can be imagined. For instance, one way to increase its linearity range is to dynamically change the dot diameter during one integration time of the sensor camera. Further studies will now investigate its practical implementation and the impact of chromaticity on wavefront sensing.

\section{Acknowledgements}
This work benefited from the support of the WOLF project ANR-$18$-CE$31$-$0018$ of the French National Research Agency (ANR). It has also been prepared as part of the activities of OPTICON H$2020$ (2017-2020) Work Package $1$ (Calibration and test tools for AO assisted E-ELT instruments). OPTICON is supported by the Horizon 2020 Framework Programme of  the  European  Commission’s  (Grant  number  $730890$). Authors are acknowledging the support by the Action Spécifique Haute Résolution Angulaire (ASHRA) of CNRS/INSU co-funded by CNES. Vincent Chambouleyron PhD is co-funded by "Région Sud" and ONERA, in collaboration with First Light Imaging. AV acknowledges funding from the European Research Council (ERC) under the European Union's Horizon 2020 research and innovation programme (grant agreement No.~757561). Finally, part of this work is supported by the LabEx FOCUS ANR-$11$-LABX-$0013$, and received the support of Action Spécifique Haute Résolution Angulaire ASHRA.


\bibliographystyle{aa} 
\bibliography{sample}

\begin{thebibliography}{17}
\expandafter\ifx\csname natexlab\endcsname\relax\def\natexlab#1{#1}\fi

\bibitem[{{Bloemhof} \& {Wallace}(2003)}]{bloemhof}
{Bloemhof}, E.~E. \& {Wallace}, J.~K. 2003, in Society of Photo-Optical
  Instrumentation Engineers (SPIE) Conference Series, Vol. 5169, Astronomical
  Adaptive Optics Systems and Applications, ed. R.~K. {Tyson} \&
  M.~{Lloyd-Hart}, 309--320

\bibitem[{Dohlen {et~al.}(2006)Dohlen, Langlois, Lanzoni, Mazzanti, Vigan,
  Montoya, Hernandez, Reyes, Surdej, \& Yaitskova}]{Kjetil2006}
Dohlen, K., Langlois, M., Lanzoni, P., {et~al.} 2006, in Ground-based and
  Airborne Telescopes, ed. L.~M. Stepp, Vol. 6267, International Society for
  Optics and Photonics (SPIE), 1093 -- 1103

\bibitem[{Fauvarque {et~al.}(2019)Fauvarque, Janin-Potiron, Correia,
  Br\^{u}l\'{e}, Neichel, Chambouleyron, Sauvage, \& Fusco}]{fauvConv}
Fauvarque, O., Janin-Potiron, P., Correia, C., {et~al.} 2019, J. Opt. Soc. Am.
  A, 36, 1241

\bibitem[{Fauvarque {et~al.}(2015)Fauvarque, Neichel, Fusco, \&
  Sauvage}]{fauvFP}
Fauvarque, O., Neichel, B., Fusco, T., \& Sauvage, J.-F. 2015, Opt. Lett., 40,
  3528

\bibitem[{Fauvarque {et~al.}(2016)Fauvarque, Neichel, Fusco, Sauvage, \&
  Girault}]{fauvOptica}
Fauvarque, O., Neichel, B., Fusco, T., Sauvage, J.-F., \& Girault, O. 2016,
  Optica, 3, 1440

\bibitem[{Guyon(2005)}]{Guyon_2005}
Guyon, O. 2005, The Astrophysical Journal, 629, 592–614

\bibitem[{Jensen-Clem {et~al.}(2012)Jensen-Clem, Wallace, \&
  Serabyn}]{jensen-clem}
Jensen-Clem, R., Wallace, J.~K., \& Serabyn, E. 2012, in 2012 IEEE Aerospace
  Conference, 1--7

\bibitem[{{N'Diaye} {et~al.}(2013){N'Diaye}, {Dohlen}, {Fusco}, \&
  {Paul}}]{ZWFSmamadou}
{N'Diaye}, M., {Dohlen}, K., {Fusco}, T., \& {Paul}, B. 2013, \aap, 555, A94

\bibitem[{N'Diaye {et~al.}(2016)N'Diaye, Vigan, Dohlen, Sauvage, Caillat,
  Costille, Girard, Beuzit, Fusco, Blanchard, Le~merrer, Le~Mignant, Madec,
  Moreaux, Mouillet, Puget, Zins, Marchetti, Close, \&
  V{\'e}ran}]{ZeldaMamadou}
N'Diaye, M., Vigan, A., Dohlen, K., {et~al.} 2016, in {SPIE Astronomical
  Telescopes + Instrumentation}, Vol. 9909, Edinburgh, United Kingdom, 99096S

\bibitem[{Ragazzoni(1996)}]{raga}
Ragazzoni, R. 1996, Journal of Modern Optics, 43, 289

\bibitem[{{Rigaut} \& {Gendron}(1992)}]{1992Rigaut}
{Rigaut}, F. \& {Gendron}, E. 1992, \aap, 261, 677

\bibitem[{Ruane {et~al.}(2020)Ruane, Wallace, Steeves, Prada, Seo, Bendek,
  Coker, Chen, Crill, Jewell, \& et~al.}]{Ruane_2020}
Ruane, G., Wallace, J.~K., Steeves, J., {et~al.} 2020, Journal of Astronomical
  Telescopes, Instruments, and Systems, 6

\bibitem[{Steeves {et~al.}(2020)Steeves, Wallace, Kettenbeil, \&
  Jewell}]{Steeves:20}
Steeves, J., Wallace, J.~K., Kettenbeil, C., \& Jewell, J. 2020, Optica, 7,
  1267

\bibitem[{{V{\'e}rinaud}(2004)}]{verinaud}
{V{\'e}rinaud}, C. 2004, Optics Communications, 233, 27

\bibitem[{{Vigan, A.} {et~al.}(2019){Vigan, A.}, {N\'{}Diaye, M.}, {Dohlen,
  K.}, {Sauvage, J.-F.}, {Milli, J.}, {Zins, G.}, {Petit, C.}, {Wahhaj, Z.},
  {Cantalloube, F.}, {Caillat, A.}, {Costille, A.}, {Le Merrer, J.}, {Carlotti,
  A.}, {Beuzit, J.-L.}, \& {Mouillet, D.}}]{viganSphere}
{Vigan, A.}, {N\'{}Diaye, M.}, {Dohlen, K.}, {et~al.} 2019, A\&A, 629, A11

\bibitem[{Wallace {et~al.}(2011)Wallace, Rao, Jensen-Clem, \&
  Serabyn}]{wallace2011}
Wallace, J.~K., Rao, S., Jensen-Clem, R.~M., \& Serabyn, G. 2011, in Optical
  Manufacturing and Testing IX, ed. J.~H. Burge, O.~W. Fähnle, \&
  R.~Williamson, Vol. 8126, International Society for Optics and Photonics
  (SPIE), 110 -- 120

\bibitem[{Zernike \& Stratton(1934)}]{zernike1934}
Zernike, F. \& Stratton, F. 1934, Monthly Notices of the Royal Astronomical
  Society, 94, 377

\end{thebibliography}

\begin{appendix}

\section{convolutional approach: general framework}
\label{appendix0}

In the infinite pupil approximation and assuming that the weighting function (energy distribution at the focal plane) $\omega$ is a real centro-symmetric function and that the focal-plane mask function $m$ is also centro-symmetric, the transfer function of a FFWFS may be written as:
\begin{equation}
\textbf{TF}=2\text{Im}\big[m\star\overline{m\omega}\big]\label{A1}\,.
\end{equation}
We consider in this formalism only pure phase mask for the Fourier filter function $m$, because the global context is the search for sensitivity of WFS so we did not  amplitude mask that would result in wasting of photons. Therefore, they are pure phase masks that can be written as $m = e^{i\Delta}$. Moreover, we continue to carry out these mathematical developments in one dimension. The previous equation \eqref{A1} becomes:
\begin{equation}
\textbf{TF}|_{f}= 2\int_{\mathbb{R}} \text{d}\text{u}~\omega|_{\text{u}}\sin\Big(
\Delta|_{\text{u}}-\Delta|_{f-\text{u}}\Big)\,, 
\label{eq:integral}
\end{equation}
where $u$ is expressed in unit of $\lambda/D$. We choose to approximate the weighting function as a rectangular and normalized function with a diameter $a$:
\begin{equation}
\omega|_u=\frac{1}{a}\mathbb{I}_{[-a/2;a/2]}\big|_u\,.
\end{equation}
The normalization allows one to respect the energy conservation. Furthermore, $a$ may be seen as the characteristic size of the energy distribution at the focal plane. $a$ is therefore the typical size of the modulation while studying the Pyramid WFS class and corresponds to the PSF characteristic length when modulation is inactive. In other words, the weighting function allows us to take into account the finite size of the pupil in spite of the "infinite pupil approximation" needed in the convolutional approach.



\section{Zernike WFS class}
\label{appendixA}

For the ZWFS, we consider the following mask: a centered dot with a diameter of $p$ and a depth $\delta$. Thus, the phase of the filtering mask equals to:
\begin{equation}
\Delta|_{\text{u}} = \delta~\mathbb{I}_{[-p/2;p/2]}\big|_u\,.
\end{equation}

\begin{figure}[htpb]
\centering
\includegraphics[scale=0.7]{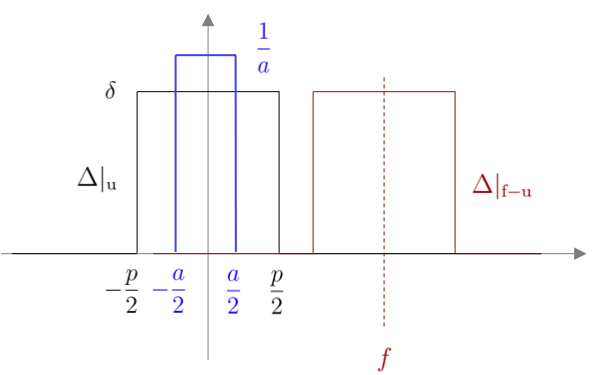}
\caption{In blue, ZWFS mask in 1 dimension with a uniform modulation representing the PSF. In black, the centered mask $\Delta|u$. In red, the shifted mask $\Delta|_{f-u}$.}
\end{figure}
The problem being symmetric, the derivation of equation \eqref{eq:integral} is only done for positive frequencies $f\geq 0$. We can distinguish two cases: whether the size of the dot is bigger than the PSF, i.e. $p\geq a$ or not, i.e. $p< a$.

\bigskip

\noindent \textbf{Case 1.} The dot is larger than the PSF: $p\geq a$.

\textbf{Case 1.1} $f\geq \frac{p+a}{2}$ \\

We have $\Delta|_{\text{u}} = \delta$ and $\Delta|_{f-\text{u}} = 0$
\begin{equation}
\textbf{TF}|_{f}= \frac{2}{a}\int_{-a/2}^{a/2} \text{d}\text{u}\sin\Big(
\Delta|_{\text{u}}-\Delta|_{f-\text{u}}\Big) = \frac{2}{a}\int_{-a/2}^{a/2}\text{d}\text{u} = 2\sin(\delta)\,.
\end{equation}

\textbf{Case 1.2} $f< \frac{p+a}{2}$

In that case, $\Delta|_{\text{u}}-\Delta|_{f-\text{u}}$ depends on $f$. We get:
\begin{eqnarray}
\textbf{TF}|_{f}&=& \text{max}\left(2\sin(\delta)-\frac{1}{a}(p+a-2f)\sin(\delta),0\right)\\&=& \text{max}\left((2f+1-\frac{p}{a})\sin(\delta),0\right)\,.
\end{eqnarray}

\noindent \textbf{Case 2.} The dot is smaller than the PSF: $p<a$.

\textbf{Case 2.1: $f\geq p$ }\\

We have again $\Delta|_{\text{u}} = \delta \ \text{and}\ \Delta|_{f-\text{u}} = 0$. Consequently, the transfer function equals to:
\begin{eqnarray}
\textbf{TF}|_{f}&=& \frac{2}{a}\int_{-a/2}^{a/2} \text{d}\text{u}\sin\Big(
\Delta|_{\text{u}}-\Delta|_{f-\text{u}}\Big)\\ &=& \frac{2}{a}\int_{-p/2}^{p/2}\text{d}\text{u}\sin(\delta) = \frac{2p}{a}\sin(\delta)\,.
\end{eqnarray}

\textbf{Case 2.2: $f< p$}

$\Delta|_{\text{u}}-\Delta|_{f-\text{u}}$ still depends on $f$. Equation \eqref{eq:integral} becomes:
\begin{eqnarray}
\textbf{TF}|_{f}&=& \frac{2}{a}\int_{-a/2}^{a/2} \text{d}\text{u}\sin\Big(
\Delta|_{\text{u}}-\Delta|_{f-\text{u}}\Big) \\ &=& \frac{2}{a}\int_{-f}^{f}\text{d}\text{u}\sin(\delta)  = \frac{2f}{a}\sin(\delta)\,.
\end{eqnarray}
The plot corresponding to these results is given earlier in the letter, Figure \ref{fig:fig1}.

\section{Pyramid WFS class}
\label{appendixB}
We derive once more the equation \eqref{eq:integral} for the PyWFS class. The phase of the filtering mask now equals to:
\begin{equation}
\Delta|_u=\alpha|u|\,,
\end{equation}
where $\alpha$ equals to $2\pi\theta/\lambda$ where $\theta$ is the pyramid apex angle and  $\lambda$ is the sensing wavelength. Equation \ref{eq:integral} then becomes:
\begin{eqnarray}
\textbf{TF}|_{f}= 2\int_{\mathbb{R}} \text{d}\text{u}~\omega|_{\text{u}}\sin\Big(\alpha(|u|-|f-u|)\Big)\,.
\end{eqnarray}

\noindent In the sine, we have:
\begin{equation}
\alpha(|u|-|f-u|) =  \left\{
\begin{array}{ll}
      -\alpha f & u\in]-\infty;0] \\
      \alpha(2u-f) & u\in]0;f[ \\
      \alpha f & u\in[f;\infty[ \\
\end{array}\,. 
\right. \label{style}
\end{equation}
Moreover $\omega$ is centro-symmetric so it is possible to reduce the integration interval:
\begin{equation}
\textbf{TF}|_{f}= 2\int_{-f}^f \text{d}\text{u}~\omega|_{\text{u}}\sin\Big(\alpha(|u|-|f-u|) \Big)\,.
\label{erer} 
\end{equation}
Figure \ref{marguerite} allows us to visualize the functions involved in previous equation. Notably, it appears that the value of the characteristic length of the weighting function over two $a/2$, i.e. the modulation radius when this device is active, plays a major role in the integration and may be seen as a cutoff frequency. Equation \eqref{erer} becomes:
\begin{figure}[ht]
\centering
\includegraphics[scale=0.7]{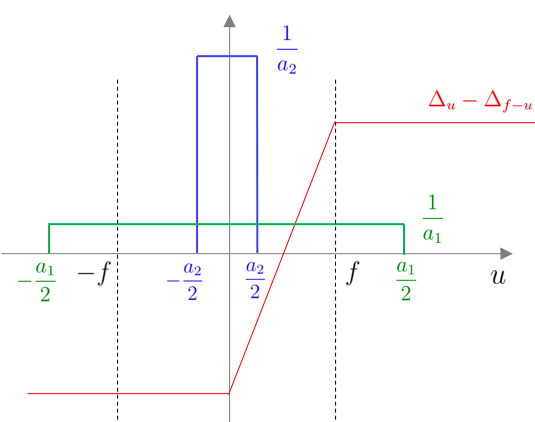}
\caption{In blue (resp. green): case of a weighting function with a small (resp. large) characteristic length. In red, function of equation \eqref{style}.}\label{marguerite}
\end{figure}
\begin{multline}
\textbf{TF}|_{f}=\frac{2}{a}\left[\int_{-\min(a/2,f)}^0 \sin(-\alpha f) \text{d}u \right.+ \\ \left.\int_{0}^{\min(a/2,f)} \sin\big(\alpha(2u-f)\big) \text{d}u\right]
\end{multline}
\vspace{-0.5cm}
\begin{multline}
=\frac{2}{a}\left[\frac{\sin(\alpha\min(a/2,f))\sin[(\min(a/2,f)-f)\alpha]}{\alpha}\right.\\-\min(a/2,f)\sin(\alpha f)\Big]\,.
\end{multline}
Finally, we get the transfer function of the Pyramid class depending on its two optical parameters, namely the Pyramid apex angle and the weighting function characteristic size.
\begin{equation}
\textbf{TF}|_{f}=  \left\{
\begin{array}{ll}
       \frac{- 2 f \sin(\alpha f)}{a} & f\leq a/2 \\
      \text{sinc}(\alpha a/2)\sin[(a/2-f)\alpha]-\sin(\alpha f) & f>a/2 \\
\end{array}\,. 
\right. \label{trtr}
\end{equation}
We may now use this formula to explain the sensitivity with respects to the spatial frequencies of the Pyramid WFS class. \\

\begin{figure}[ht]
\centering
\includegraphics[scale=0.5]{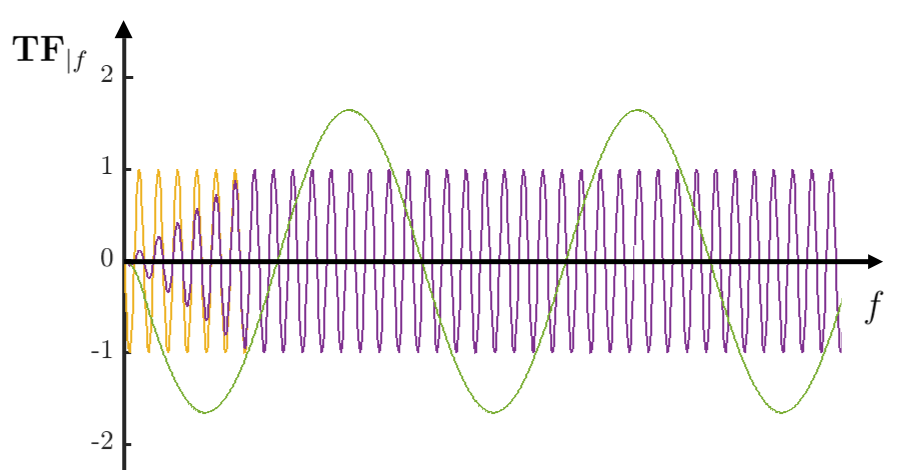}
\caption{Graphical visualisation of transfer function, i.e. Equation \eqref{trtr}, for the PyWFS configurations of Figure \ref{fig:comparisonFourier}: non-modulated PyWFS in yellow, modulated PyWFS in purple and FPyWFS in green.}\label{melanie}
\end{figure}

Firstly, we are interested in apex angle parameter influence on the sensitivity. In other words we study the difference between classical and flattened pyramids. To do so, we assume that the modulation is inactive. Consequently, the parameter $a$ is related to the PSF size. Considering a pupil diameter of D, a sensing wavelength of $\lambda$ and a imaging focal of $f_{\text{oc}}$, we have:
\begin{equation}
a=\frac{\lambda f_{\text{oc}}}{D}\,.    
\end{equation}
To distinguish between flattened or classical pyramids, we just have to identify the limit apex angle $\theta_{\text{limit}}$ allowing to totally separate pupil images. It can actually be linked with the pupil diameter and the imaging system focal via the following formula:
\begin{equation}
\theta_{\text{limit}}=\frac{D}{f_{\text{oc}}}\,.
\end{equation}
Consequently, if $\theta$ is below $\theta_{\text{limit}}$ there is an overlap of the pupil images and the Pyramid is therefore a flattened one whereas a $\theta$ above $\theta_{\text{limit}}$ implies a complete separation of the pupil images and thus, a classical pyramid. If we use the $a$ and $\alpha$ variables, these two cases may be summarized in the following way:
\begin{eqnarray}
a\alpha < \pi && \text{Flattened Pyramid}\\
a\alpha \geq \pi && \text{Classical Pyramid}
\end{eqnarray}
Such a distinction allows to explain why the sensitivity of the flattened pyramid may be larger in absolute value than the classical pyramid one. As a matter of fact, the sinc function in equation \eqref{trtr} may be significant for some frequencies when $a\alpha$ is small, i.e. for flattened pyramids. Keeping in mind that the sensitivity is linked with the \textbf{TF} via the convolution product of Equation \eqref{eq:sens}, this shows why the FPyWFS sensitivity may reach 2 when the classical pyramid optimally attains 1.  Moreover, the small $\alpha$ implies large oscillations with respects to the spatial frequencies (see green curve of Figure \ref{melanie}). This is relevant regarding to the observed sensitivity which indeed oscillates (dotted green curve in Figure \ref{fig:comparisonFourier}). 

By contrast, if $a\alpha$ is large, i.e. if pupil images are completely separated, the sinc function may be neglected and the transfer function can be summarized as:
\begin{equation}
\textbf{TF}|_{f}=  \left\{
\begin{array}{ll}
       \frac{- 2f \sin(\alpha f)}{a} & f\leq a/2 \\
      -\sin(\alpha f) & f>a/2 \\
\end{array}\,. 
\right. \label{trtr2}
\end{equation}
This function corresponds may be seen as the transfer function of the classical pyramid. We notice that it oscillates more rapidly than the flattened pyramid one (yellow and purple curves of Figure \eqref{melanie}). However, these oscillations disappear when we look at the corresponding sensitivity curves (Figure \eqref{fig:comparisonFourier}). Such a peculiarity may be explained by Equation \eqref{eq:sens}: to get the sensitivity curve, the transfer function is convoluted with the PSF which is in this case larger than oscillation period. As a result, the transfer function is smoothed and the sensitivity follows the \textbf{TF} envelope.

Concerning this envelope, we may observe two regimes. The first one goes from the null spatial frequency to the modulation radius $a/2$ ; it is linear with $f$. The second one corresponds to spatial frequencies above the modulation radius ; it is constant and equals to 1. We identify here the typical behavior of the classical pyramid (modulated or not) with its two regimes, slope and phase sensors separated by a cutoff frequency corresponding to the modulation radius \citep{verinaud}.

The convolutional approach therefore shows its capability to describe the sensitivity of pyramid sensors. It indeed allows to get a unique formula which explains both the enhanced and oscillating sensitivity of the FPyWFS and the dual behavior slope/phase sensors of the classical modulated PyWFS.

\end{appendix}

\end{document}